\documentclass[english,aps , prl , reprint  ,inputenc=utf8,superscriptaddress]{revtex4-1}
\usepackage[T1]{fontenc}
\usepackage[utf8]{inputenc}
\usepackage{color}
\usepackage{babel}
\usepackage{booktabs}
\usepackage{amstext}
\usepackage{graphicx}
\usepackage[unicode=true,pdfusetitle,
 bookmarks=true,bookmarksnumbered=false,bookmarksopen=false,
 breaklinks=false,pdfborder={0 0 1},backref=false,colorlinks=false]
 {hyperref}
\hypersetup{
 colorlinks=true,citecolor=blue,linkcolor=blue,urlcolor=blue}

\makeatletter

\providecommand{\tabularnewline}{\\}

\makeatother

\begin{document}
\title{Pressure Correction for Solvation Theories}
\author{Anton Robert}
\affiliation{PASTEUR, Département de chimie, École normale supérieure, PSL University,
Sorbonne Université, CNRS, 75005 Paris, France}
\author{Sohvi Luukkonen}
\affiliation{Maison de la Simulation, USR 3441 CNRS-CEA-Université Paris-Saclay,
91191 Gif-sur-Yvette, France}
\author{Maximilien Levesque}
\email{maximilien.levesque@ens.fr}
\affiliation{PASTEUR, Département de chimie, École normale supérieure, PSL University,
Sorbonne Université, CNRS, 75005 Paris, France}
\begin{abstract}
Liquid state theories such as integral equations and classical density
functional theory often overestimate the bulk pressure of fluids because
they require closure relations or truncations of functionals. Consequently,
the cost to create a molecular cavity in the fluid is no longer negligible
and those theories predict wrong solvation free energies. We show
how to correct them simply by computing an optimized Van der Walls
volume of the solute and removing the undue free energy to create
such volume in the fluid. Given this versatile correction, we demonstrate
that state-of-the-art solvation theories can predict, within seconds,
hydration free energies of a benchmark of small neutral drug-like
molecules with the same accuracy as day-long molecular simulations.
\end{abstract}
\maketitle
The ability to predict accurately solvation free energies (SFEs) and
solvent maps unlocks the access to several key thermodynamical observables
of biomolecular systems \citep{klimovich_predicting_2010} like relative
solubilities, binding free energies \citep{snyder_mechanism_2011,wang_ligand_2011},
transfer free energies \citep{moeser_role_2015} or partition coefficients
\citep{bannan_calculating_2016}. SFEs can be rigorously computed
with methods involving molecular simulations and free energy perturbation
(FEP) techniques \citep{zwanzig_hightemperature_1954,shirts_2008}.
Those are time and resource consuming: they require tens to thousands
of CPU hours on high-performance computers. Implicit-solvent models
\citep{tomasi_medium_nodate} that ignore the molecular description
of the solvent were designed in order to cope the expensive cost of
FEP, but solving Poisson equation in a dielectric continuum often
delivers inaccurate results \citep{mobley_small_2009,fennell_oil/water_2010,wagoner_solvation_2004}
and/or require quantum mechanical calculations \citep{klamt_cosmo-rs_2016}.
 Recently, considering the necessity of evaluating precisely solvation
free energies in the drug design process, major actors of the pharmaceutical
drug discovery industry publicly called the academic world for alternatives,
pointing out the lack of precision of current methods or their high
numerical cost \citep{sherborne_collaborating_2016}.

An alternative lies within liquid state theories \citep{hansen_theory_2013}.
Indeed, solvation theories like the molecular density functional theory
(MDFT) \citep{jeanmairet_molecular_2013} or the three-dimensional
reference interaction site model (3D-RISM) integral equation \citep{beglov_integral_1997},
are now able to predict hydration free energies of complex solutes
like proteins \citep{ding_efficient_2017,imai_solvation_2004} or
aluminosilicate surfaces \citep{levesque_solvation_2012}. At their
roots, they solve the molecular Ornstein-Zernike (MOZ) equation using
two different approaches in order to compute estimations of SFEs in
few minutes at most. MDFT, for instance, minimizes a free energy functional
$\mathcal{F}[\rho(\mathbf{r},\omega)]$ of a six-dimensional solvent
molecular density $\rho(\mathbf{r},\omega)$, where $\omega=(\theta,\phi,\psi)$
are the three Euler angles of the rigid solvent molecule at position
$\mathbf{r}$ relative to a frozen three-dimensional solute. The free
energy minimization happens in the external potential generated by
each atom of the solute modeled, for instance, by the same Lennard-Jones
potentials and point charges as in a molecular dynamics simulation.
The minimum of the free energy functional is the solvation free energy
of the given solute. The density that minimizes the functional is
the equilibrium spatial and angular map of solvent molecules $\rho_{\text{eq}}(\mathbf{r},\omega)$,
often described as the molecular solvent map. Since the functional
is unknown, one most-often truncates it to a density expansion at
second order, which can be shown to be equivalent to the well-known
hyper-netted chain (HNC) approximation in the integral equation theory.
A MDFT minimization in the HNC approximation is 3 to 5 orders of magnitude
faster than FEP/alchemical methods relying on molecular dynamics (MD)
in predicting SFEs \citep{ding_efficient_2017}. The speed-up increases
with the size of the solute.

Liquid state theories in the HNC and other approximations overestimate
the pressure of the bulk solvent \citep{evans_failure_1983,rickayzen_integral_1984,powles_density_1988}.
The HNC bulk pressure of most common models of water like TIP3P \citep{TIP3P}
or SPCE \citep{berendsen_missing_1987_SPCE} at 300K and 1 kg per
liter is around $10000$ atm instead of 1 atm. Since the free energy
to create a molecular cavity within a fluid increases with its volume
and the pressure in the fluid, those theories that overestimate the
pressure also overestimate the SFE. For small molecules, SFEs predicted
in the HNC approximation are not even in qualitative agreement with
experiments, for which the precision is about half a kcal/mol on modern
calorimetric apparatus.

To assess the accuracy of MDFT and elaborate the pressure correction
(PC), we use the FreeSolv database \citep{duarte_ramos_matos_approaches_2017}.
It contains experimental and predicted \footnote{Predictions were computed with alchemical transformation from flexible
solute MD simulation with TIP3P\citep{TIP3P} parameters for the solvent
and GAFF (v.1.7) \citep{wang_development_2004} parameters with AM1-BCC
partial charges \citep{jakalian_fast_2000,jakalian_fast_2002} for
the solutes. The same force field parameters were used for MDFT and
MC calculation in this paper.} hydration free energies of 642 small neutral drug-like molecules.
Since MDFT computes the SFE of rigid solutes \footnote{The effect of a molecule flexibility is in principle not an obstacle
for MDFT. Several conformers can be identified and their SFEs contribute
to a weighted average. This work is ongoing.}, we did reference calculations for rigid solutes by using Hybrid-4D
MC simulations \citep{belloni_2019}. Briefly summarized, Belloni's
Hybrid-4D computation of solvation free energy is based on two parallel
out-of-equilibrium Monte Carlo (MC) simulations, one of bulk water
and one of the solvated solute, and the Jarinsky equality \citep{jarzynski_nonequilibrium_1997}.
In the bulk water (resp. solvated) simulation, the rigid solute is
slowly inserted (resp. deleted) via a fictive 4th dimension every
100 iterations. The work of each insertion/deletion is calculated
and the Bennett acceptance ration \citep{bennett_efficient_1976}
is used to combine the insertion and deletion distributions to predict
the SFE.

\begin{figure}
\noindent \begin{centering}
\includegraphics[width=8.5cm]{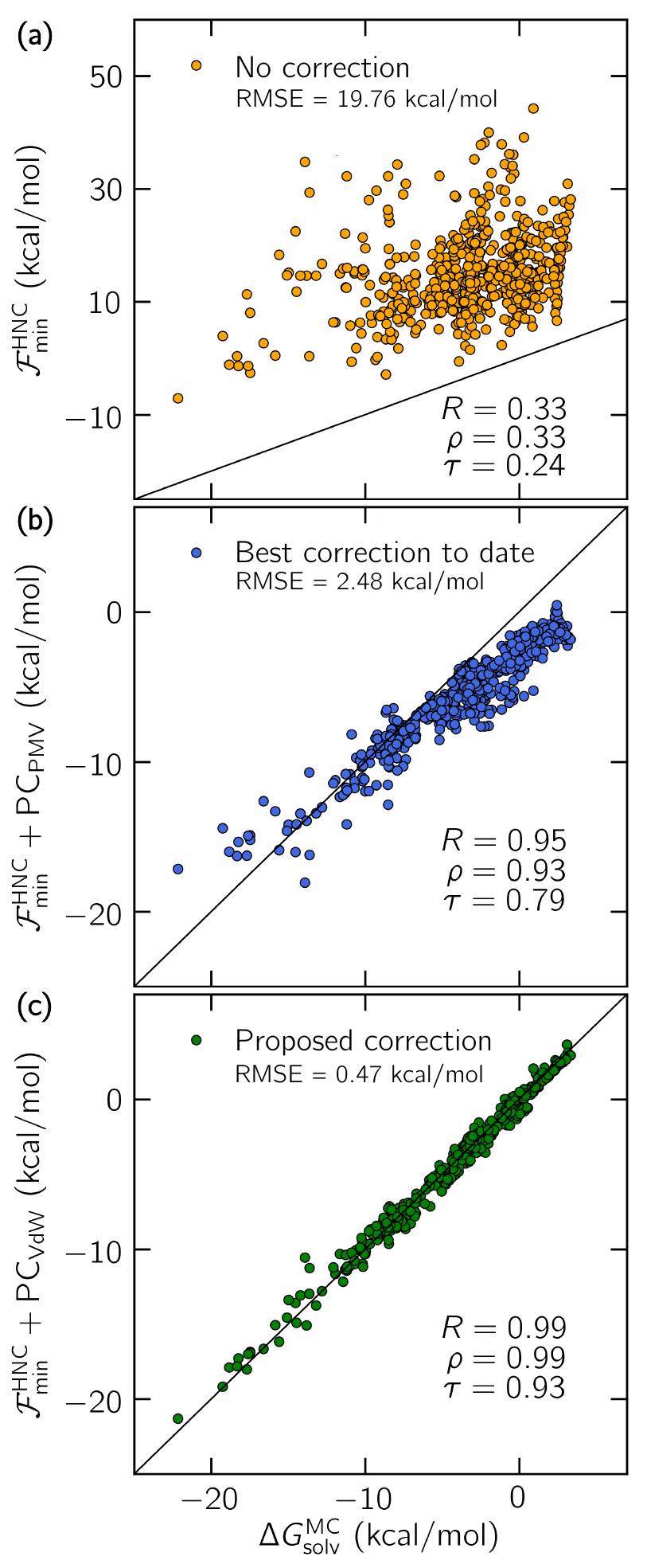}
\par\end{centering}
\caption{Comparison between the hydration free energies of the FreeSolv database
(642 neutral drug-like molecules) of reference simulations on the
horizontal axis and of MDFT-HNC (a) without correction, (b) with the
$\text{PC}_{\text{PMV}}$ correction \citep{sergiievskyi_fast_2014},
and (c) the $\text{PC}_{\text{VdW}}$ correction of this work. The
statistical error bars of the simulations (3 standard deviations)
are smaller than the size of the dots. The statistical measures at
the bottom-right corner refer to Pearson's $R$, Spearsman's $\rho$
and Kendall's $\tau$. \label{fig:Comparison-of-two}}
\end{figure}

In Fig. \ref{fig:Comparison-of-two}.a, we compare SFEs approximated
by MDFT-HNC \footnote{The MDFT calculations were done with a cubic supecell of $24$ Å with
a spatial resolution of 0.33 Å and an angular resolution of 84 orientations
per spatial grid node. We report 31 non-converging molecules so that
all statistical measures in this article are given with respect to
the 611 molecules that converged.} and calculated by the abovementionned state-of-the-art reference
MC calculations. The root mean squared error (RMSE) is 19.8 kcal/mol.
Clearly, HNC SFEs must be corrected. In the case of MDFT-HNC with
a pressure-based correction (PC), it reads

\begin{equation}
\Delta G_{\text{solv}}=\min_{\rho}(\mathcal{F}[\rho(\mathbf{r},\omega)])+\text{PC}.\label{eq:solv-1}
\end{equation}

Several paths have already been followed to pressure-correct SFEs:
from empirical fits of the error on experimental values \citep{palmer_towards_2010,ratkova_accurate_2010,ratkova_solvation_2015,roy_performance_2019},
to semi-empirical corrections without parameters like the partial
molar volume correction proposed by Sergiievskyi et al. \citep{sergiievskyi_fast_2014,sergiievskyi_pressure_2015}.
The latter correction is based on the idea that at the macroscopic
scale, the free energy to create a cavity of volume $V$ in a fluid
of pressure $P$ is $PV$. Thus, in the macroscopic limit, if the
pressure is $P_{\text{HNC}}=10000$~atm instead of $P_{\text{Exp}}=1$
atm, the pressure correction (PC in Eq.$\ $\ref{eq:solv-1}) is $\text{PC}=-\left(P_{\text{HNC}}-P_{\text{Exp}}\right)V$.
Even if this correction is justified in the macroscopic limit, it
is not at the molecular scale. If one uses the unambiguous partial
molar volume (PMV) noted $\Delta V$ as the molecular volume, one
comes back to Sergiievskyi's proposition. The PMV can be derived rigorously
in liquid state theories inherently in the grand canonical ensemble
like MDFT from the variation $\Delta N$ in the number of solvent
molecules in the MDFT supercell while inserting the solute at constant
temperature, volume and solvent chemical potential. The PMV pressure
correction is thus $\text{PC}_{\text{PMV}}=-\Delta P\Delta V=(P_{\text{HNC}}-P_{\text{Exp}})\Delta N/n_{b}$
where $n_{b}$ is the bulk solvent density. As shown in Fig.$\ $
\ref{fig:Comparison-of-two}b, the PMV pressure correction improves
drastically the predicted solvation free energies \citep{sergiievskyi_fast_2014,sergiievskyi_pressure_2015,misin_communication:_2015,chong_thermodynamic-ensemble_2015},
yielding a root mean square error to reference simulations of 2.48
kcal/mol compared to 19.76 kcal/mol for the uncorrected ($\text{PC}=0$
in Eq.$\ $\ref{eq:solv-1}) HNC results.

We now present the volume optimized pressure correction that uses
a geometrical definition of the molecular volume, that is the volume
of overlapping Van der Waals (VdW) spheres centered on every atom
of the solute. The radii depend upon the chemical nature of each atom
and were initially taken from \citep{bondi_van_1964} that gathers
multiple experimental estimations. Since those 10 radii are not unambiguously
defined and subject to large incertitude, we optimized them by about
6\% in average around Bondi's experimental values so that the $PV$
pressure correction minimizes the RMSE of MDFT compared to reference
calculations. The VdW volumes were iteratively calculated and optimize
via the Nelder-Mead algorithm \citep{gao_implementing_2012,ajd98_utilities_2019}
using a bootstrap technique on a subset of 288 molecules from the
FreeSolv database.

\paragraph*{MDFT vs MC}

We first optimize the VdW radii on reference SFEs calculated by molecular
simulations. In order to discard all possible error compensation effects
due to force field and flexibility, we evaluated the performances
of MDFT with respect to MC simulations on rigid molecules. Thus, we
first minimize the RMSE of MDFT with respect to the rigid MC simulations.
The optimized radii are reported in Table \ref{tab:Optimized-VdW-radius}.
The comparison between MDFT SFE predictions with PC$_{\text{VdW}}$
and MC on the whole FreeSolv database is shown in Fig.$\ $\ref{fig:Comparison-of-two}c.
The proposed VdW pressure correction divides the average error by
a factor of 5 with respect to the MC simulations: the RMSE is now
$0.47$ kcal/mol. Correlations are also improved with $R=0.99$ and
Kendall's $\tau=0.93$ with PC$_{\text{VdW}}$ compared to $R=0.95$
and $\tau=0.79$ with PC$_{\text{PMV}}$ . Altough the optimization
of the 10 radii was conducted on less than half the molecules in the
database, those results obtained on the whole database show a high
transferability to the other molecules\emph{.}

\begin{table*}[t]
\noindent \begin{centering}
\begin{tabular}{ccccccccccc}
\toprule 
{\footnotesize{}VdW radius (\r{A})} & {\footnotesize{}C} & {\footnotesize{}N} & {\footnotesize{}O} & {\footnotesize{}H} & {\footnotesize{}F} & {\footnotesize{}Cl} & {\footnotesize{}Br} & {\footnotesize{}I} & {\footnotesize{}P} & {\footnotesize{}S}\tabularnewline
\midrule
\midrule 
{\footnotesize{}Initial values \citep{bondi_van_1964}} & {\footnotesize{}1.70} & {\footnotesize{}1.55} & {\footnotesize{}1.52} & {\footnotesize{}1.20} & {\footnotesize{}1.47} & {\footnotesize{}1.75} & {\footnotesize{}1.85} & {\footnotesize{}1.98} & {\footnotesize{}1.80} & {\footnotesize{}1.80}\tabularnewline
\midrule 
{\footnotesize{}Optimized vs. Sim.} & {\footnotesize{}1.711} & {\footnotesize{}1.734} & {\footnotesize{}1.588} & {\footnotesize{}1.318} & {\footnotesize{}1.59} & {\footnotesize{}1.815} & {\footnotesize{}1.872} & {\footnotesize{}1.982} & {\footnotesize{}1.458} & {\footnotesize{}1.721}\tabularnewline
\midrule 
{\footnotesize{}Optimized vs. Exp.} & {\footnotesize{}1.682} & {\footnotesize{}1.893} & {\footnotesize{}1.430} & {\footnotesize{}1.353} & {\footnotesize{}1.510} & {\footnotesize{}1.887} & {\footnotesize{}1.984} & {\footnotesize{}1.960} & {\footnotesize{}1.426} & {\footnotesize{}1.804}\tabularnewline
\bottomrule
\end{tabular}
\par\end{centering}
\caption{Van der Walls radii used for the optimized pressure-correction $\text{PC}_{\text{VdW}}$
in MDFT. First raw: initial values as taken for experiments. Second
and third raw: optimized versus reference FEP SFE calculations or
experimental SFEs. \label{tab:Optimized-VdW-radius}}
\end{table*}

\paragraph*{MDFT vs Experiment}

\begin{figure}
\begin{centering}
\includegraphics[width=8.5cm]{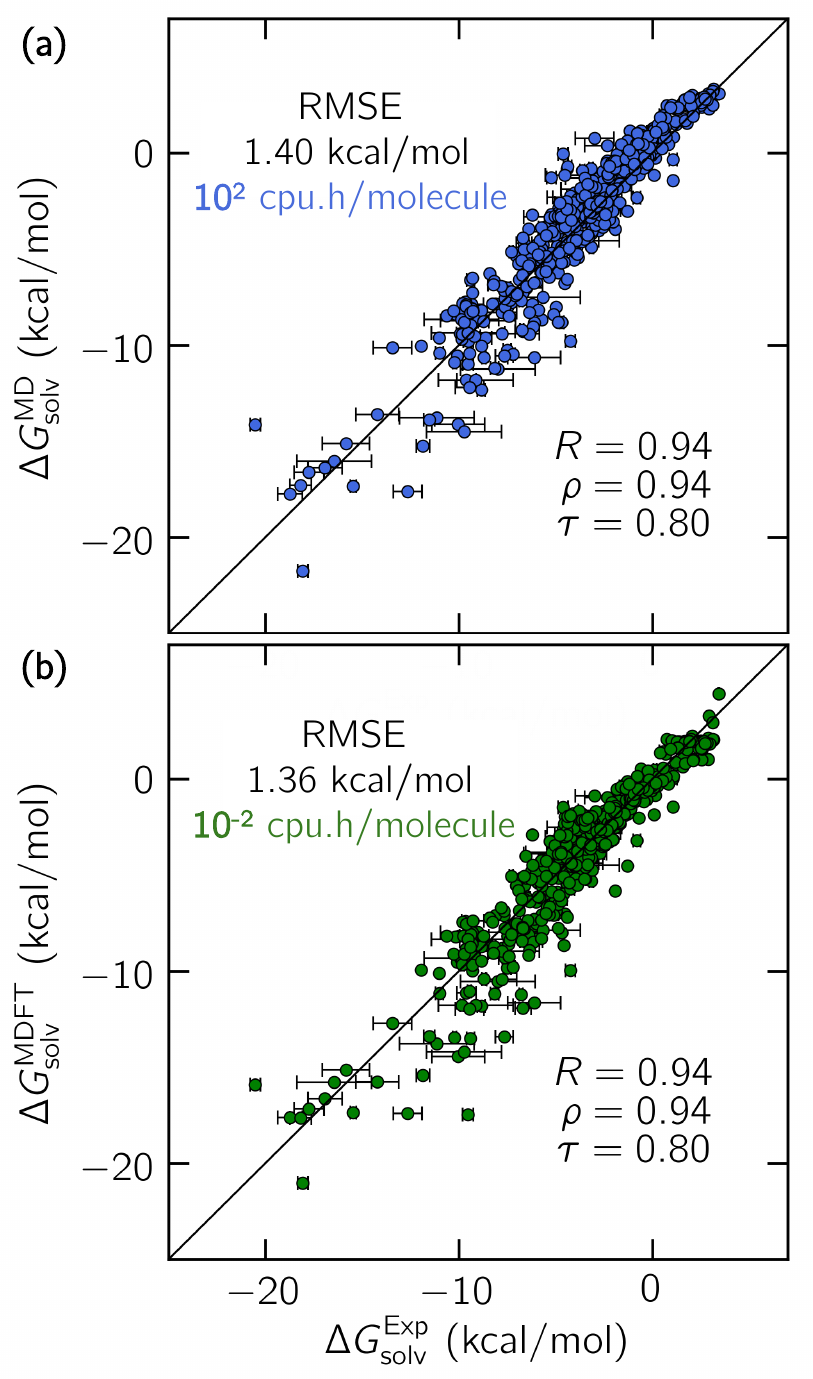}
\par\end{centering}
\caption{(a) Reference free energy calculations based on MD and (b) MDFT predictions
of hydration free energies compared to experimental values. MDFT is
corrected using the optimized volume pressure correction $\text{PC}_{\text{VdW}}$
as described in the text. Experimental uncertainties are reported
for 31\% of the FreeSolv molecules: error bars (of lenght the provided
uncertainty) are drawn, with an average 0.47 kcal/mol. The statistical
measures at the bottom-right corner refer to Pearson's $R$, Spearsman's
$\rho$ and Kendall's $\tau$. \label{fig:exp}}
\end{figure}

We now turn to optimizing the VdW radii on experimental SFEs. Since
MDFT computes the SFE of rigid solutes, we restrict ourselves to rigid
molecules. To this purpose, we compute the deviation between SFEs
we obtained with single conformer MC simulations and SFEs values computed
via alchemical transformation from flexible solute MD simulation given
in the FreeSolv database. If the difference in SFE of the rigid conformer
and of the flexible molecule is below $0.1$ kcal/mol, the molecule
is considered rigid. That is the case of 288 molecules among the 642
of the Freesolv database, those used in the paragraph above for consistency.
Then, we optimize the VdW radii with respect to the experimental SFE
of those 288 ``rigid'' molecules. The final VdW raddi are reported
in Table \ref{tab:Optimized-VdW-radius}. In Fig.$\ $\ref{fig:exp}b,
we show the comparison between PC$_{\text{VdW}}$-corrected MDFT-HNC
SFEs and the experimental SFEs for the whole dataset of molecules,
including those that are flexible. The RMSE of MDFT compared to the
experiment is $1.36$ kcal/mol, thus reaching the same accuracy as
reference FEP simulations with respect to experiments : the RMSE between
FEP SFEs and experiments is 1.40 kcal/mol (see Fig.\ref{fig:exp}a).
Note that each MDFT's SFE prediction takes few seconds to compute
on a 8 cores-laptop \citep{ding_efficient_2017}.

The pressure correction we introduced in this paper is simple, versatile
and efficient. Using this correction, we have compared MDFT, a state-of-the-art
solvation theory, with experimental and simulation results to assess
its capability to predict solvation free energies. In order to discard
all possible error compensation effects due to force field and flexibility,
for instance, we evaluated the performances of MDFT with respect to
MC simulations. MDFT can predict SFEs of small drug-like molecules
with the same accuracy\textcolor{magenta}{{} }as MC simulations (RMSE
of $0.47$ kcal/mol for MDFT vs MC), in few seconds on a laptop. Optimizing
the pressure correction on experimentally measured SFE of rigid molecules,
we reached the same accuracy as flexible MD state-of-the-art simulations
coupled with FEP (RMSE of $1.36$ kcal/mol for MDFT vs Exp). This
Van der Waals pressure correction can be applied to any liquid state
theory that overestimates the pressure of the bulk fluid, like 3D-RISM
for instance.

This paper shows that solvation free energies and thus affinities
(or binding free energies) can be predicted four orders of magnitude
faster with the molecular density functional theory than with state-of-the-art
molecular simulations methods, \emph{without} trading off accuracy.
In the context of \emph{in silico} drug discovery, this means that
screening accurately chemical libraries containing millions of molecules
is now possible in the time scale of days.

\bibliographystyle{achemso}
\bibliography{main}

\end{document}